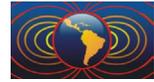

# VARIACION DIURNA ESTACIONAL DEL CAMPO GEOMAGNETICO REGISTRADO EN EL OBSERVATORIO DE HUANCAYO

Domingo Rosales[1*], Erick Vidal[1]

[1] Observatorio Geomagnético de Huancayo – Instituto Geofísico del Perú, Huancayo, Perú.
*e-mail: domingo.igp@gmail.com.

**ABSTRACT**

In this article we study the seasonal effect on the diurnal variation of the geomagnetic field registered in the Huancayo Observatory, located in the Magnetic Equator, which is driven by "ionospheric currents" and its counterpart induced by "telluric currents". Huancayo Observatory has the highest amplitude in the diurnal variation, because of being in the Magnetic Equator and under the "Equatorial Electrojet". We present the pattern of seasonal variation in diurnal variation of components X, Y and Z, the same as confirmed by previous works since 1940. The effect of solar activity cycle of about 11 years in the diurnal variation is also confirmed; it is observed that amplitudes are greater in the maximum of solar activity.

**RESUMEN**

En el presente artículo se estudia el efecto estacional en la variación diurna del campo geomagnético registrado en el observatorio de Huancayo, localizado en zona del ecuador magnético que es gobernada por las "corrientes ionosféricas" y su contraparte inducida por las "corrientes telúricas". El Observatorio de Huancayo al estar en la zona del ecuador magnético registra la mayor amplitud en la variación diurna, esto es por el hecho de encontrarse en la zona Magnética Ecuatorial y debajo del "Electrochorro Ecuatorial". En el presente artículo se presenta el patrón de la variación estacional en la variación diurna de las componentes X, Y y Z, las mismas que confirman con trabajos previos realizados desde el año 1940. El efecto del ciclo de actividad solar aproximado de 11 años en la variación diurna también es confirmado, se observa que las amplitudes son mayores en los máximos de actividad solar.

**Palabras Clave**: Variación diurna, variación estacional, geomagnetismo, actividad solar.

**Introducción**

El campo geomagnético que es registrado en un observatorio es el resultado de la contribución de varias fuentes. Para un mejor estudio se ha clasificado las fuentes del campo magnético de la Tierra en fuentes internas y externas. En el observatorio de Huancayo, 99% del campo magnético que se registra proviene de fuentes internas (principalmente del núcleo externo) y aproximadamente un 1% proviene de fuentes externas (principalmente debido al efecto solar). El campo geomagnético cambia en escalas de tiempo que van desde milisegundos a millones de años, siendo las fuentes externas las que poseen escalas de tiempo mas cortas, y pueden ser separadas de las contribuciones de las fuentes internas para su estudio.

Las variaciones del campo geomagnético debido a las fuentes externas proporcionan información de la actividad externa, fundamentalmente del Sol. Algunos de los efectos de la actividad solar son: la variación diurna, variación estacional, periodo fundamental de rotación solar de 27.0 días, ciclo de actividad solar de 11 y 22 años, tormentas geomagnéticas, efecto Forbush, etc. A esto se tiene que agregar el efecto de traslación de la Tierra alrededor del Sol (lo que contribuye en una variación del tipo estacional), y la rotación de la Tierra (variación diurna).

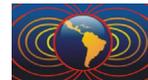

La variación diurna se produce por el efecto combinado de los sistemas de corrientes ionosféricas que fluyen en la región E y su correspondiente corriente inducida que fluye bajo la superficie terrestre denominado "corrientes telúricas" creado por efecto dinamo. Este sistema consiste de dos vórtices, uno en cada hemisferio con focos aproximadamente a 30º de latitud y localizado en el tiempo aproximadamente una hora antes del medio día local. Ambos vórtices comprenden dos sistemas de corrientes, uno fluye por encima de la superficie de la Tierra (dirección Oeste-Este) y el otro fluye bajo la superficie de la Tierra (en dirección opuesta). Ambos sistemas de corrientes son fijos en espacio y tiempo en relación con el Sol, fluyendo en todo momento, la Tierra gira bajo ello dando lugar a la variación regular diurna denominada "variación $S_R$". La intensidad de las corrientes y las amplitudes de los vórtices cambian a lo largo del año manifestándose en la variación estacional, siendo mayores en la época de verano para cada hemisferio (Svalgaard, 2007). El ciclo de actividad solar de ~11 años también manifiesta su efecto en la variación diurna, por lo que la amplitud de la variación estacional es mayor en el máximo de actividad solar y menor en un mínimo de actividad solar.

El observatorio geomagnético de Huancayo (lat. 12.04º S, long. 75.33º O, alt. 3314 m.s.n.m), se encuentra localizado dentro de la zona del ecuador magnético (año 2015.0: lat. mag. 2.28º S, long. mag. 2.65º O, Dip -0.24º). Los primeros registros de la variación diurna en el observatorio de Huancayo han sido realizados desde el año 1919 cuando se instalo una estación magnética en las proximidades del hoy observatorio de Huancayo. Ya en aquellas épocas se observó una anormalmente gran amplitud en la variación diurna de la componente H, que era de 3 a 5 veces mayor que las variaciones registradas en otros puntos. Los primeros estudios detallados de este efecto fueron realizados por Bartels y Johnston (1940a; 1940b) y Egedal (1947), quienes determinaron que el rango diurno de la componente H en estaciones ecuatoriales próximas al ecuador magnético presentan picos y que ésta es particularmente mayor en el observatorio de Huancayo con un rango de 125 nT. El mecanismo que produce esta amplia variación fue propuesto como una banda de ~300 km de ancho que fluye sobre el ecuador magnético a una altitud aproximada de 100 km, que más adelante fue denominado por Chapman (1951) "electrochorro ecuatorial" (EEJ). Forbush y Casaverde (1961) estudiaron las características del EEJ en las componentes H, D y Z en la zona del ecuador magnético y asumieron que el EEJ no afecta o es muy poca su contribución en el cambio de la variación diurna de la componente D. Sin embargo nuevos trabajos como los de Rastogi (1996; 1998) y Okeke (1998) han demostrado que existen variaciones estacionales en la componente D (Okeke and Hamano, 2000). Por lo tanto el patrón de la variación estacional anual puede ser atribuido a la variabilidad de los procesos ionosféricos y estructura física tales como la conductividad y la estructura de los vientos, que son los responsables de la variación diurna.

Rastogi (1998) afirma que Huancayo es la única estación ecuatorial magnética que registra el incremento simultáneo de la componente X y Y en horas que corresponden al periodo de día (07:00 – 11:00 LT) y que cuando X decrece (12:00 – 18:00 LT), o se hace negativo, también la componente Y hace lo mismo. Por lo tanto la variación diurna de Y en latitudes ecuatoriales sugiere ser una parte constitutiva del sistema del EEJ.

El observatorio de Huancayo por a su ubicación geográfica, además de ser la estación que registra una de las mayores amplitudes diurnas en su componente H (o X) debido al efecto ionosférico, también registra un adicional incremento debido al efecto propio del EEJ.

**Datos**

Son usados los datos promedios horarios de los elementos geomagnéticos X, Y y Z del observatorio de Huancayo desde 1998.0 al 2015.0 (total 447048 datos). Los datos de los años 2000, 2008 y 2014 que corresponden a años de máximos y mínimos de actividad solar de los ciclos solares 23 y 24, son tomados en consideración para verificar el efecto del ciclo de actividad solar en la variación diurna.

**Análisis**

La variación del campo geomagnético debió a las fuentes internas (variación secular), fueron extraídas quedando solo las variaciones debido a fuentes externas. Los datos son agrupados de modo tal que permita mostrar la variación diurna mensual y son ordenados mes a mes para su análisis correspondiente.

**Resultados y discusión**

La variación diurna cambia día a día en las tres componentes. Un análisis anual muestra que existe un patrón en el comportamiento que varia mes a mes. Tomando como referencia el hemisferio sur (por la ubicación del observatorio de Huancayo), para la componente X, la máxima amplitud se presenta en la zona de los equinoccios (marzo-abril y setiembre-octubre), y la mínima amplitud se produce en la zona de los solsticios (diciembre-enero y junio-julio). Para la componente Y, la máxima amplitud se presenta en la zona del solsticio de verano (diciembre-enero), y la mínima amplitud se registra en los meses de abril y setiembre. En la componente Z, la máxima amplitud se registra en la zona del solsticio de verano (diciembre-enero), y la mínima en la zona del solsticio de invierno (junio-julio). Los rangos máximo de la variación diurna son 210, 55 y 35 nT para las componentes X, Y y Z respectivamente (Fig. 1). Estas características guardan concordancia y reconfirman los resultados obtenidos por Okeke y Hamano (2000), Rosales y Vidal (2013; 2015), y trabajos previos en cuyos casos para el análisis de la variación estacional se utilizaron la clasificación mediante los denominados "meses de Lloyd" (D: *noviembre, diciembre, enero y febrero*, E: m*arzo, abril, setiembre y octubre*, J: m*ayo, junio, julio y agosto*).

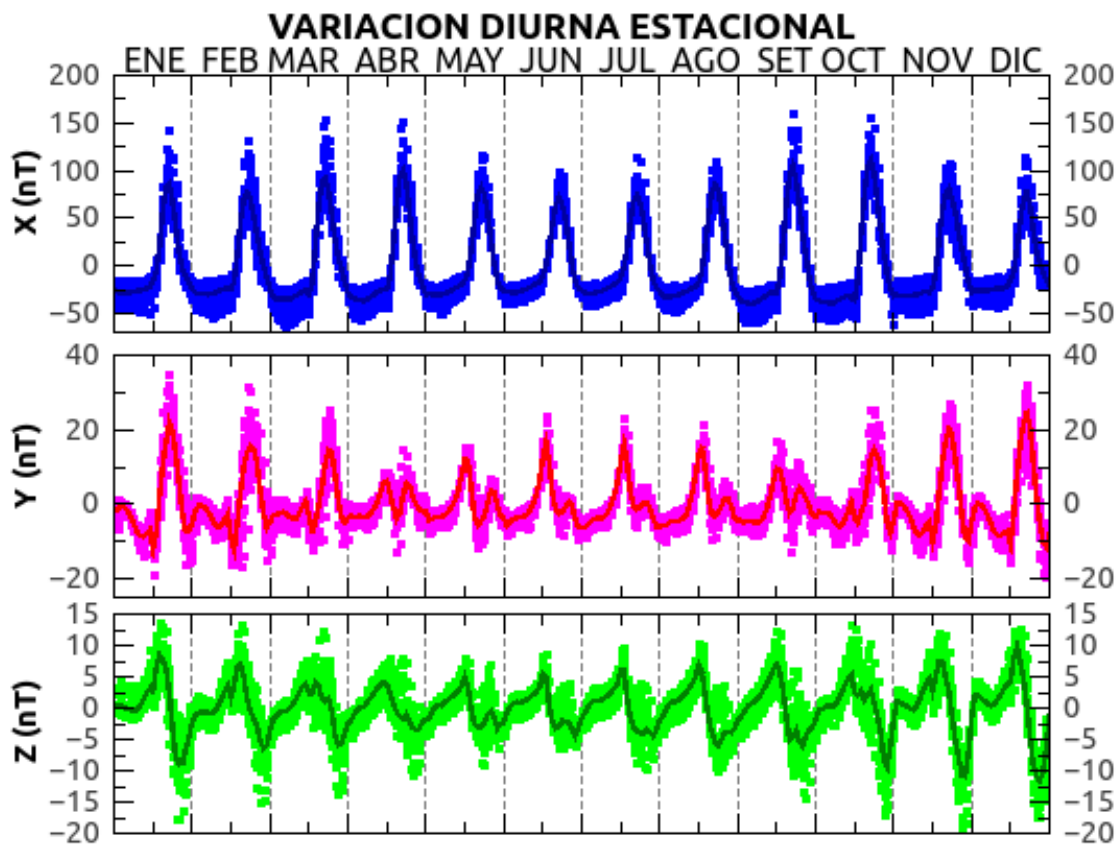

**Figura 1**. Variación diurna estacional 1998.0-2015.0

Por otro lado, M. Abbas (Abbas *et al.*, 2013), realiza evaluaciones de variación estacional a datos de promedios horarios de los 5 días más quietos internacionales por mes del año 2008, de la estación magnética de Ancón (Lat. 11.77 S, Long. 77.15 O) mediante la clasificación de los meses de Lloyd, concluyendo que la máxima amplitud en la componente H se registra en el mes Lloyd-J. Esta conclusión no guarda relación con el presente trabajo y muchos otros trabajos realizados previamente. Tomando en cuenta que el observatorio geomagnético de Huancayo y la estación de Ancón estan próximos, las características entre ambos puntos deben ser bastante similares, por lo que se recomienda realizar un nuevo análisis en los datos de Ancón.

El ciclo de actividad solar de ~11 años también actúa en la variación diurna y en la variación estacional. En un año de mínimo de actividad solar la variación diurna de las amplitudes también son mínimas, y en años que corresponde a máximos de actividad solar las amplitudes de las variaciones diurnas son mayores. Así, el máximo de actividad solar del ciclo solar 23 correspondió al año 2000 y se verifico una mayor amplitud en las variaciones diurnas (Fig. 2, color verde). El máximo de actividad solar del ciclo solar 24 se produjo el año 2014, manifestando una mayor amplitud en las variaciones diurnas pero menor a lo producido en el año 2000 (Fig. 2, color azul). El mínimo de actividad solar del ciclo solar 23 correspondió al año 2008, donde se verifica una menor amplitud en las variaciones diurnas (Fig. 2, color rojo).

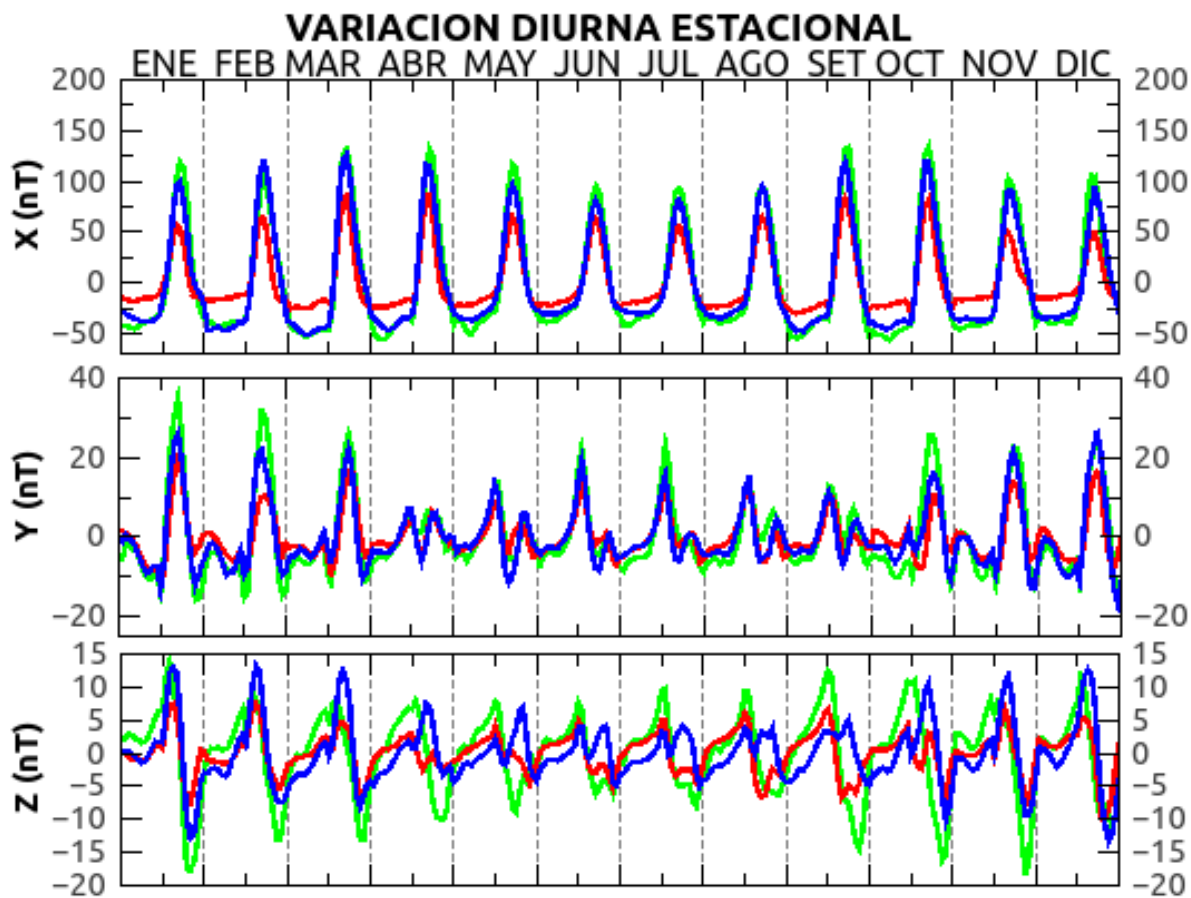

**Figura 2**. Variación diurna estacional. Máximo de actividad ciclo solar 23 (año 2000 en color verde). Mínimo de actividad ciclo solar 23 (año 2008 en color rojo). Máximo de actividad ciclo solar 24 (año 2014 en color azul)

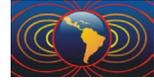

Se observa que durante seis meses (desde octubre hasta marzo) el incremento es simultaneo en las componentes X y Y, en horas que corresponde a la mañana (07:00 – 11:00 LT) y posterior reducción en su intensidad por la tarde (12:00 -18:00 LT), y durante los seis meses siguientes (desde abril hasta setiembre) la variación en la componente Y es inverso disminuyendo por las mañanas y subiendo por las tardes (Figs. 1 y 2), lo que demostraría que las corrientes meridional y zonal forman parte del sistema EEJ.

**Conclusiones**

El patrón de la variación estacional en las tres componentes X, Y y Z son consistentes con trabajos previos. El efecto de variación estacional influye en la variación día a día sobre las tres componentes; esta variabilidad observada sugiere que existen cambios estacionales en la zona de la ionósfera como son: viento y conductividad. La variación estacional observada con máxima amplitud en X en los equinoccios puede ser por una mayor densidad de electrones y vientos presente en los equinoccios.

Las conclusiones sobre la variación estacional de Ancón realizadas por M. Abbas (Abbas *et al.*, 2013) no son concordantes con el presente trabajo ni otros trabajos previos, sobre todo tomando en consideración que Huancayo y Ancón son estaciones magnéticas muy próximas, por lo que ambas puntos deben registrar características similares, por lo que se recomienda realizar un nuevo análisis en los datos de Ancón del año 2008.

La clasificación en meses de Lloyd permite mostrar el efecto de la variación estacional en la variación diurna, sin embargo un análisis del tipo mes por mes muestra una mejor resolución del efecto de la variación estacional. La forma de onda de la variación diurna en la componente X es similar en los 12 meses, en tanto que en las componente Y y Z, cambian mes a mes. En las componentes Y y Z las amplitudes son máximas en verano, y son menores o mínimas en invierno guardando plena concordancia con los estudios realizados por Svalgaard (2007).

El ciclo de actividad solar de ~11 años también manifiesta su efecto en las variaciones diurnas sobre las tres componentes, siendo mayor su amplitud en los máximos (años 2000 y 2014) y menor amplitud en los mínimos (año 2008).

**Referencias**